\documentclass[aps,prd,eqsecnum,preprint,tightenlines,nofootinbib,showpacs]{revtex4-1}
\usepackage{graphicx,amsmath,latexsym}

\newcommand{\bea}{\begin{eqnarray}}
\newcommand{\eea}{\end{eqnarray}}
\newcommand{\be}{\begin{equation}}
\newcommand{\ee}{\end{equation}}
\newcommand{\ub}[1]{\underline{#1}}
\newcommand{\ob}[1]{\overline{#1}}
\newcommand{\Pminus}{{\cal P}^-}

\begin{document}

\title{A nonperturbative coupled-cluster method for quantum field theories\footnote{Presented
at CIPANP 2012, the Eleventh Conference on the Intersections of Particle and Nuclear Physics,
May 28 - June 3, 2012, St.\ Petersburg, Florida.}}

\author{J. R. Hiller}
\affiliation{Department of Physics \\
University of Minnesota-Duluth \\
Duluth, Minnesota 55812}

\date{\today}

\begin{abstract}
The nonperturbative Hamiltonian eigenvalue problem for bound states
of a quantum field theory is formulated in terms of Dirac's light-front
coordinates and then approximated by the exponential-operator technique 
of the many-body coupled-cluster method.  This approximation eliminates
any need for the usual approximation of Fock-space truncation.  
Instead, the exponentiated operator is truncated, and the terms retained
are determined by a set of nonlinear integral equations.  These
equations are solved simultaneously with an effective eigenvalue
problem in the valence sector, where the number of constituents
is small.  Matrix elements can be calculated, with extensions of
techniques from many-body coupled-cluster theory, to obtain form
factors and other observables.
\end{abstract}


\maketitle


\section{Introduction}

The nonperturbative solution of quantum field theories
is a highly nontrivial task and yet is crucial for
the full understanding of a strongly coupled theory
such as quantum chromodynamics.  Lattice methods
have been quite successful in attacking this problem
but suffer from the limitation to a Euclidean 
formulation and the lack of direct access to
wave functions.  Light-front Hamiltonian methods~\cite{Dirac,DLCQreviews}
are not as well developed, but they do remain in Minkowski
space and hold the promise of obtaining wave functions,
as coefficients in Fock-state expansions of the 
Hamiltonian eigenstates.

The Hamiltonian eigenvalue problem is typically
solved by first truncating Fock space.  This yields
a finite set of coupled equations for a finite
collection of wave functions.  However, the
truncation induces various complications, not
the least of which are uncanceled divergences
and Fock-sector dependencies~\cite{SecDep}.
The light-front coupled-cluster (LFCC) method~\cite{LFCClett}
avoids these complications by not truncating
Fock space and instead approximating the
relationship between wave functions to obtain
a finite set of equations for a still infinite
collection of wave functions.  There is a price
to be paid, of course; the resulting equations
are nonlinear and the calculation of matrix
elements requires particular care.  The latter
is greatly aided by the close similarity of
the mathematics (though not the physics) of
the many-body coupled-cluster method~\cite{CCorigin}
for nonrelativistic nuclear physics and 
quantum chemistry~\cite{CCreviews}.

\section{Overview of the method}

The LFCC method~\cite{LFCClett} is designed to approximate the
light-front Hamiltonian eigenvalue problem
$P^+\Pminus|\psi\rangle=(M^2+P_\perp^2)|\psi\rangle$
by writing the eigenstate as $|\psi\rangle=\sqrt{Z}e^T|\phi\rangle$.
Here $|\phi\rangle$ is a valence state with a fixed, small
number of constituents; $T$ is an operator that increases
particle number and conserves all relevant quantum
numbers; and $\sqrt{Z}$ is a normalization factor.
The approximation is that $T$ is truncated to one or
a few operators but the Fock space used as a basis
for the eigenstate is not truncated.  As a consequence
of the approximation, wave functions associated with
higher Fock states are restricted to what $T$ can
generate from the lower-state wave functions.

To implement this approach, we define an effective
Hamiltonian $\ob{\Pminus}=e^{-T}\Pminus e^T$ and
a projection $P_v$ onto the valence sector, with
$1-P_v$ restricted to projection onto just enough 
sectors to fully constrain the form of $T$.  The
resulting LFCC equations are
\be
P_v\ob{\Pminus}|\phi\rangle=\frac{M^2+P_\perp^2}{P^+}|\phi\rangle, \;\;\;\;
(1-P_v)\ob{\Pminus}|\phi\rangle=0.
\ee
The second equation is essentially an auxiliary equation
for $T$, which is needed to define the effective Hamiltonian $\ob{\Pminus}$
used in the first, a valence eigenvalue equation.

Matrix elements are computed from the right and left eigenstates
of $\ob{\Pminus}$.  For example, consider the 
expectation value for an operator $\hat{O}$:
\be
\langle\hat O\rangle=\frac{\langle\phi| e^{T^\dagger}\hat O e^T|\phi\rangle}
                      {\langle\phi| e^{T^\dagger} e^T|\phi\rangle}.
\ee
Direct computation requires an infinite sum over Fock space, which is
intractable.  Instead, we can borrow some mathematics from the
many-body coupled-cluster method~\cite{CCreviews}.  We define
$\ob{O}=e^{-T}\hat O e^T$ and 
$\langle\widetilde\psi|=\langle\phi|\frac{e^{T^\dagger}e^T}
      {\langle\phi|e^{T^\dagger} e^T|\phi\rangle}$.  The
expectation value can then be expressed as
$\langle\hat O\rangle=\langle\widetilde\psi|\ob{O}|\phi\rangle$, and
the ket $\langle\widetilde\psi|$ is normalized as
\be
\langle\widetilde\psi'|\phi\rangle
=\langle\phi'|\frac{e^{T^\dagger}e^T}{\langle\phi| e^{T^\dagger} e^T|\phi\rangle}|\phi\rangle
=\delta(\ub{P}'-\ub{P}).
\ee
The effective operator $\ob{O}$ can be computed from its
Baker--Hausdorff expansion,
$\ob{O}=\hat O + [\hat O,T]+\frac12[[\hat O,T],T]+\cdots$.
The ket $\langle\widetilde\psi|$ is a left eigenvector of $\ob{\Pminus}$,
as can be seen from
\be
\langle\widetilde\psi|\ob{\Pminus}
=\langle\phi|\frac{e^{T^\dagger}\Pminus e^T}{\langle\phi| e^{T^\dagger} e^T|\phi\rangle}
=\langle\phi|\ob{\Pminus}^\dagger \frac{e^{T^\dagger}e^T}
                            {\langle\phi| e^{T^\dagger} e^T|\phi\rangle}
=\frac{M^2+P_\perp^2}{P^+}\langle\widetilde\psi|.
\ee
Physical quantities can then be computed from the LFCC eigenstates.

This approach has been applied to a light-front analog of the Greenberg--Schweber 
model~\cite{GreenbergSchweber,LFCClett}, where the
lowest-order $T$ produces the exact answer, and to QED~\cite{LFCCqed}, in
a calculation of the dressed-electron state and its anomalous magnetic moment.
In addition, a technique for including zero modes has been developed~\cite{LFCCzeromodes}
and applied to $\phi^3$ and $\phi^4$ theories and to the Wick--Cutkosky model.
Much of this work is less briefly summarized in various
proceedings~\cite{LC2011,QNP2012,QCDatWork2012}.

\section{Summary}

The LFCC method provides a nonperturbative Hamiltonian approach to
the solution of quantum field theories that avoids Fock-space truncation
and the consequent difficulties of uncanceled divergences and sector
dependence.  It can be applied to any regulated light-front Hamiltonian
and is systematically improvable through the addition of
terms to the $T$ operator, classified by the net increase in particle
number and the number of annihilation operators.  Zero modes can be
included, to allow consideration of theories with symmetry breaking.

Work on QED continues, with consideration of additional terms
to include electron-positron pairs and of the bound states of
muonium and positronium.
Many interesting applications remain to be considered, with the
goal of understanding the method well enough to apply it to QCD.
Light-front holographic QCD~\cite{hQCD} may be a useful starting point.


\acknowledgments
This work was done in collaboration with S.S. Chabysheva
and supported in part by the US Department of Energy.

\end{document}